\documentclass[aps,pre,twocolumn,superscriptaddress,preprintnumbers,showpacs,nofootinbib]{revtex4-1}

\usepackage{amsmath,amssymb}
\usepackage[dvipdf,dvips,dvipdfmx]{graphicx}
\usepackage{here}
\usepackage{color}

\newcommand{\Tr}{\mathrm{Tr}}

\begin{document}

\title{Cost Reduction of Swapping Bonds Part in Anisotropic Tensor Renormalization Group}

\author{Hideaki Oba}
\email[]{h\_oba@hep.s.kanazawa-u.ac.jp}

\affiliation{Institute for Theoretical Physics, Kanazawa University,
 Kanazawa 920-1192, Japan}

\date{\today}

\begin{abstract}
The bottleneck part of anisotropic tensor renormalization group (ATRG) is a swapping bonds part
which consists of a contraction of two tensors 
and a partial singular value decomposition of
a matrix,
and their computational costs are $O(\chi^{2d+1})$,
where $\chi$ is the maximum bond dimension and $d$ is the dimensionality of a system.
We propose an alternative method for the swapping bonds part 
and it scales with $O(\chi^{\max(d+3,7)})$, 
though the total cost of 
ATRG with the method
remains $O(\chi^{2d+1})$.
Moreover, the memory cost of the whole algorithm
can be reduced from $O(\chi^{2d})$ to $O(\chi^{\max(d+1,6)})$. 
We examine ATRG with or without 
the proposed method in the four-dimensional Ising model and find that
the free energy density of the proposed algorithm is consistent with
that of the original ATRG 
while the elapsed time is significantly reduced.
We also compare the proposed algorithm with higher-order tensor renromalization group (HOTRG) 
and find that the value of the free energy density of the proposed algorithm is lower than that of HOTRG
in the fixed elapsed time.
\end{abstract}

\preprint{KANAZAWA-19-05}


\maketitle

\section{Introduction}
Tensor network algorithms
are theoretically and numerically
useful tools for
quantum and classical many-body systems \cite{Intro_TN_1,Intro_TN_2}.
For example,
to study a one-dimensional quantum many-body system,
one can use density matrix renormalization group \cite{DMRG,DMRG_PRB,DMRGreview},
which is based on matrix product states \cite{MPS_1,MPS_2}
and can be understood as a variational method for a wave function.
For higher dimensional cases, 
there are algorithms based on
projected entangled pair states \cite{PEPS,PEPS_2,PEPS_3}.
In addition to those algorithms,
tensor renormalization group (TRG) \cite{TRG}
is an approach for coarse-graining tensor networks
and calculates
the partition function
of a two dimensional classical model.
For studying two or higher dimensional classical systems,
we can use higher-order tensor renormalization group (HOTRG) \cite{HOTRG}.
There are series of
the improved algorithms for the coarse graining
on the basis of various philosophies
\cite{SRG,TEFR,TNR,TNR_PRB,loopTNR,hamiltonian_TRG,RSVDTRG,Gilt,EBO,FET,PTTRG,BTRG,ATRG}.

Since the tensor network algorithms avoid the sign problem,
they have received attention not only from condensed matter physics but also from high energy physics
\cite{
Banuls:2013jaa,Kuhn:2015zqa,Banuls:2015sta,Banuls:2016lkq,Banuls:2016gid,Banuls:2017ena,
Shimizu:2012zza,
Shimizu:2014uva,
Shimizu:2014fsa,
Shimizu:2017onf,
Denbleyker:2013bea,Yu:2013sbi,Meurice:2016mkb,
Takeda:2014vwa,Kawauchi:2016xng,Sakai:2017jwp,Kadoh:2018hqq}.
We hope that the lattice models suffering from the sign problem in high energy physics,
(such as quantum chromodynamics with finite quark density, the $\theta$ vacuum of quantum chromodynamics, chiral gauge theory and the super-symmetric model)
can be analyzed by the tensor network algorithms.
However, if one computes physical quantities of such high dimensional systems with the algorithms,
the computational costs and the memory costs get worse and 
it is difficult for the algorithms to 
provide results
with sufficient precision.

A novel algorithm, anisotropic tensor renormalization group (ATRG), has been proposed recently \cite{ATRG}.
It is one of the tensor network algorithms for coarse-graining tensor networks
and can be applied to an arbitrary dimensional lattice model like HOTRG.
Moreover, it takes much lower computational-cost than HOTRG.
The bottleneck part in ATRG 
is a swapping bonds part 
which consists of a contraction of two tensors 
and a partial singular value decomposition (PSVD) of
a matrix.
The computational cost and the memory cost 
of the part are $O(\chi^{2d+1})$ and $O(\chi^{2d})$ respectively,
where $\chi$ is the maximum bond dimension
and $d$ is the dimensionality of a system.
If the costs of the swapping bonds part in ATRG are reduced,
we can expect to obtain more accurate results
than those from the original ATRG
with the same elapsed time and the same memory resource.

In this paper, we propose a swapping bonds method
of which 
the computational cost can be reduced from $O(\chi^{2d+1})$ to $O(\chi^{\max(d+3,7)})$
and the memory cost of the whole algorithm can be reduced from $O(\chi^{2d})$ to $O(\chi^{\max(d+1,6)})$.
Furthermore, we numerically examine the accuracy of the free energy density of 
the four-dimensional Ising model and
the elapsed time
of ATRG using the proposed method
to compare with the original ATRG and HOTRG.

The rest of this paper is organized as follows.
In Sec.~\ref{sec:ATRG}, 
we review the original ATRG algorithm for a four-dimensional hypercubic system.
In Sec.~\ref{sec:cr_swapping}, 
the proposed swapping bonds method is explained.
In Sec.~\ref{sec:numerical_results},
we present numerical results of 
the four-dimensional Ising model
for a comparison between ATRG with the proposed method and the original ATRG or HOTRG
and discuss their performances.
A summary and outlooks are given in Sec.~\ref{sec:summary}.
In appendix~\ref{sec:making_projectors}, we explain
a method for making tensors, $E$ and $F$, in ATRG.
In appendix~\ref{sec:rsvd}, we
study the parameters of the randomized singular value decomposition (RSVD).

\section{Anisotropic Tensor Renormalization Group \label{sec:ATRG}}
\begin{figure}[t]
\centering
\includegraphics[width=85mm]{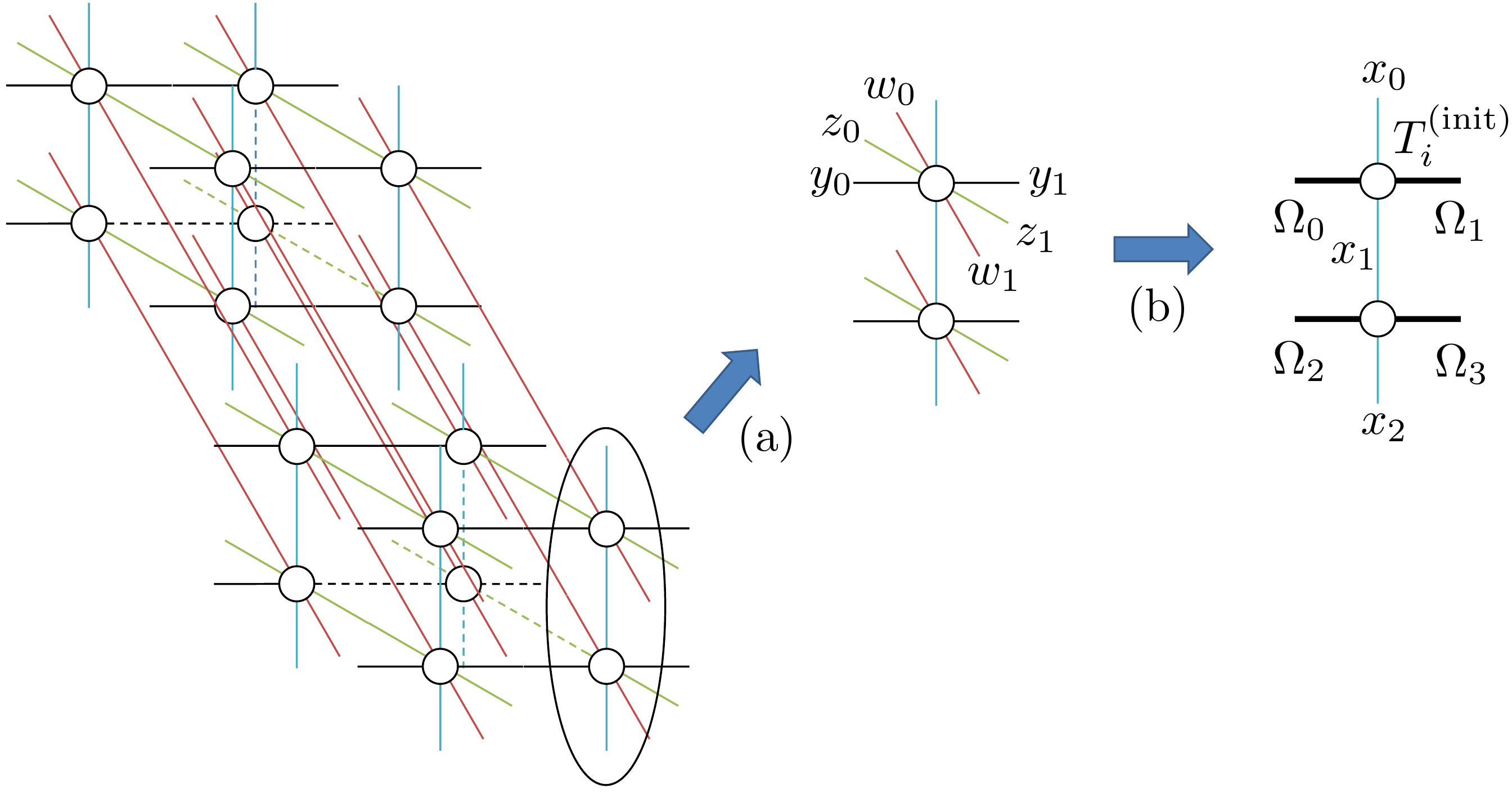}
 \caption{The initial tensor $T^{(\mathrm{init})}_i$ in 
   the initial tensor network of a four-dimensional system. 
 (a) We focus on two tensors in the initial tensor network.
 (b) If the direction of the coarse graining
   is set in $x$-axis, 
   we redefine the initial tensor as
   $T^{(\mathrm{init})}_{i;\Omega_0 \Omega_1 x_0x_1}$,
   where $\Omega_0=(y_0,z_0,w_0)$ 
   and $\Omega_1=(y_1,z_1,w_1)$ are bundles of 
   the bond indices.
 }
 \label{fig:4d_initial_tensor}
\end{figure}

In this section, 
we explain the original ATRG \cite{ATRG} for a four-dimensional hypercubic system.
First of all, 
we rewrite the partition function of the system in terms of a tensor network as
\begin{align}
  Z = \Tr \prod_i T^{(\mathrm{init})}_{i;x_0x_1y_0y_1z_0z_1w_0w_1},
\end{align}
where $T^{(\mathrm{init})}_i$ is an initial tensor,
$i$ runs all lattice sites,
and $\Tr$ means summing up all the bond indices of the initial tensors.
If the direction of the coarse graining
is set in $x$-axis
\footnote{
  Another axis, such as $y,z$ or $w$, can be 
  treated in a similar way.
},
which is the vertical direction 
in Fig.~\ref{fig:4d_initial_tensor},
we redefine the initial tensor as
$T^{(\mathrm{init})}_{i;\Omega_0 \Omega_1 x_0x_1}$,
where $\Omega_0=(y_0,z_0,w_0)$ 
and $\Omega_1=(y_1,z_1,w_1)$ are bundles of 
the bond indices
\footnote{
  $\Omega_0$ and $\Omega_1$ 
  consist of the combination of $y_0,z_0,w_0$ and 
  that of $y_1,z_1,w_1$ respectively.
  For example,
  $\Omega_0=y_0+(z_0-1)\chi^{(\mathrm{init})}+(w_0-1)(\chi^{(\mathrm{init})})^2$
  and $\Omega_1=y_1+(z_1-1)\chi^{(\mathrm{init})}+(w_1-1)(\chi^{(\mathrm{init})})^2$
  if all of the bond indices run from 1 to $\chi^{(\mathrm{init})}$.
  Although the bond dimensions of the each tensor legs are different in some case,
  we can make the bundles similarly.}.

\begin{figure}[t]
  \centering
  \includegraphics[width=85mm]{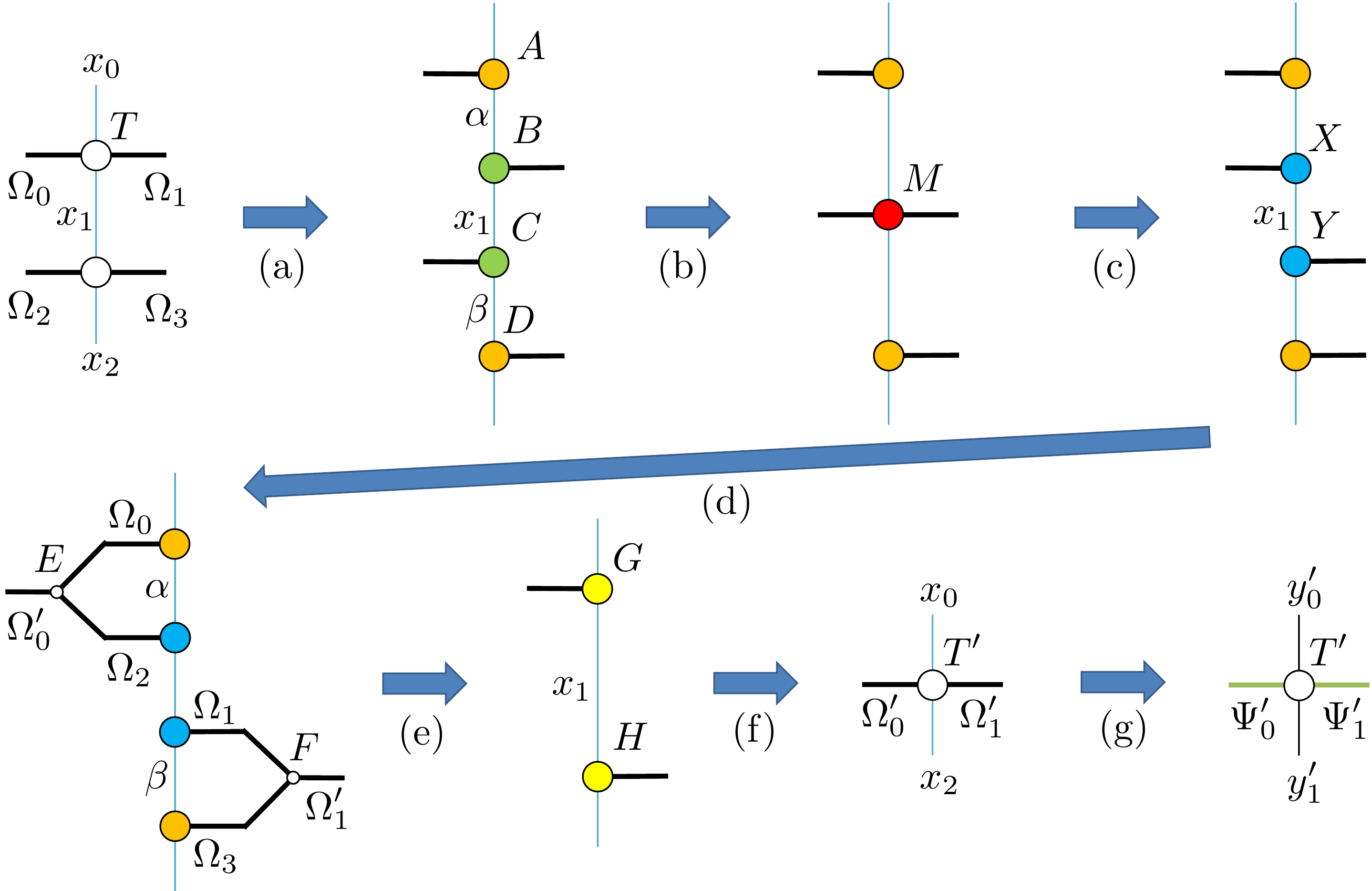}
  \caption{The flow of the original ATRG.
   In step (a), a tensor $T$ is decomposed into
   two tensors such as $A$ and $B$ or $C$ and $D$ with the SVD.
   In step (b), the two tensors, $B$ and $C$, are contracted by summing up the bond $x_1$ 
   and a tensor $M$ is obtained.
   In step (c), $M$ is decomposed by the PSVD into two tensors, $X$ and $Y$,
   to change
   the combinations of bonds from $\Omega_1 \alpha$ of $B$ and $\Omega_2\beta$ of $C$ to 
   $\Omega_2 \alpha$ of $X$ and $\Omega_1 \beta$ of $Y$.
   In step (d), two tensors,
   $E$ and $F$, are made from $A,X,Y$ and $D$ with a method
   is explained in appendix~\ref{sec:making_projectors}.
   In step (e), we contract $A,X$ and $E$ by summing up the bonds $\alpha,\Omega_0,\Omega_2$ to make $G$,
   and contract $D,Y$ and $F$ by summing up the bonds $\beta,\Omega_1,\Omega_3$ to make $H$.
   In step (f), $G$ and $H$ are contracted by summing up the bond $x_1$ to obtain a coarse-grained tensor $T'$.
   In step (g), we reorder the bonds of $T'$
   for the next direction of the coarse graining, $y$-axis.
 }
 \label{fig:atrg_algorithm}
\end{figure}

Then, we execute the coarse graining of the tensor network to reduce the degrees of freedom using ATRG \cite{ATRG}.
The whole flow of ATRG is shown in Fig.~\ref{fig:atrg_algorithm} 
where there are seven steps, (a)--(g).
First, in step (a), a tensor $T$
\footnote{$T$ is the initial tensor at the first coarse graining.}
is decomposed  
with the singular value decomposition (SVD),
\begin{align}
  T_{\Omega_0 \Omega_1 x_0x_1}\simeq \sum_{\alpha=1}^\chi U_{\{T\} \Omega_0 x_0\alpha}S_{\{T\} \alpha\alpha}V_{\{T\} \Omega_1 x_1\alpha},
\end{align}
where $\alpha$ is a new bond for $x$-axis.
Then, we make four tensors, $A,B,C$ and $D$, as
\begin{align}
  A_{\Omega_0 x_0\alpha}&=U_{\{T\}\Omega_0 x_0\alpha},\\
  B_{\Omega_1 x_1\alpha}&=S_{\{T\}\alpha\alpha}V_{\{T\}\Omega_1 x_1\alpha},\\
  C_{\Omega_0 x_0\alpha}&=S_{\{T\}\alpha\alpha}U_{\{T\}\Omega_0 x_0\alpha},\\
  D_{\Omega_1 x_1\alpha}&=V_{\{T\} \Omega_1 x_1\alpha}.
\end{align}
In step (b), the two tensors, $B$ and $C$, are contracted and a tensor $M$ is obtained by
\begin{align}
  M_{\Omega_1 \Omega_2 \alpha \beta } = \sum_{x_1} B_{\Omega_1 x_1\alpha}C_{\Omega_2 x_1\beta}.
\end{align}
Then in step (c), we decompose $M$ into two tensors, $X$ and $Y$, to change
the combinations of the bonds from 
$\Omega_1 \alpha$ of $B$ and $\Omega_2\beta$ of $C$ to 
$\Omega_2 \alpha$ of $X$ and $\Omega_1 \beta$ of $Y$ as
\begin{align}
  M_{\Omega_1 \Omega_2 \alpha \beta }&\simeq \sum_{x_1=1}^\chi U_{\{M\}\Omega_2 \alpha x_1}S_{\{M\}x_1x_1}V_{\{M\}\Omega_1 \beta x_1},\label{eq:dec_M}\\
  X_{\Omega_2 \alpha x_1} &= \sqrt{S_{\{M\}x_1x_1}}U_{\{M\}\Omega_2 \alpha x_1},\\
  Y_{\Omega_1 \beta x_1} &= \sqrt{S_{\{M\}x_1x_1}}V_{\{M\}\Omega_1 \beta x_1}.
\end{align}
One may use the PSVD, such as the Arnoldi method \cite{Arnoldi} and the RSVD \cite{RSVD}, for the decomposition of $M$ in eq. (\ref{eq:dec_M}).
In step (d), two tensors,
$E$ and $F$, are made from $A,X,Y$ and $D$ with a method
explained in appendix~\ref{sec:making_projectors}.
In step (e), we contract $A,X$ and $E$ to make $G$,
and contract $D,Y$ and $F$ to make $H$ as
\begin{align}
  G_{\Omega'_0 x_0x_1} &= \sum_{\alpha,\Omega_0,\Omega_2} A_{\Omega_0 x_0\alpha}X_{\Omega_2 \alpha x_1}E_{\Omega_0 \Omega_2 \Omega'_0},\\
  H_{\Omega'_1 x_1x_2} &= \sum_{\beta,\Omega_1,\Omega_3} D_{\Omega_3 x_2\beta}Y_{\Omega_1 \beta x_1}F_{\Omega_1 \Omega_3 \Omega'_1},
\end{align}
where $\Omega'_0=(y'_0,z'_0,w'_0)$ and $\Omega'_1=(y'_1,z'_1,w'_1)$.
Then, in step (f), $G$ and $H$ are contracted to obtain a coarse-grained tensor $T'$,
\begin{align}
  T'_{\Omega'_0 \Omega'_1 x_0x_2} = \sum_{x_1} G_{\Omega'_0 x_0x_1} H_{\Omega'_1 x_1x_2}.
\end{align}
Finally, in step (g), we reorder the bonds of $T'$
for the next direction of the coarse graining, $y$-axis,
and redefine $T'$ as $T'_{\Psi'_0 \Psi'_1 y'_0y'_1}$
in the same way as the initial tensor,
where $\Psi'_0=(x_0,z'_0,w'_0)$ and $\Psi'_1=(x_2,z'_1,w'_1)$.

\begin{figure}[t]
  \centering
  \includegraphics[width=85mm]{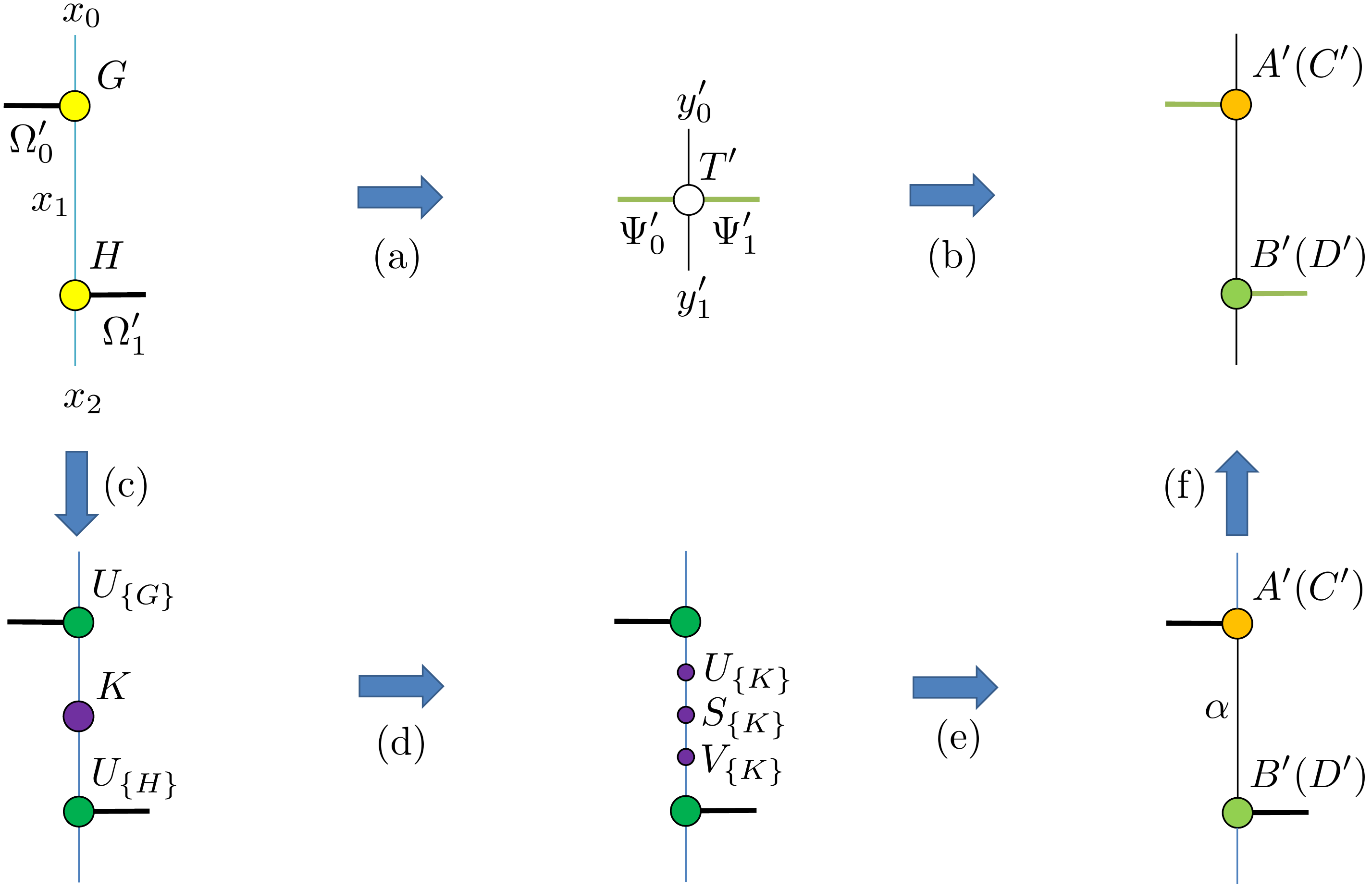}
  \caption{
   The flow of the improved method to make $A',B',C'$ and $D'$ from $G$ and $H$ \cite{ATRG}.
   Step (a) and (b) represent the step (f), (g) and (a) in Fig.~\ref{fig:atrg_algorithm}.
   In step (c), $G$ and $H$ are decomposed in the coarse-grained direction to obtain $K$ 
   by the contraction in eq. (\ref{eq:make_K}).
   In step (d), we execute the SVD of $K$.
   In step (e), $A',B',C'$ and $D'$ are made by the contraction in eq. (\ref{eq:A_dash})--(\ref{eq:D_dash}).
   In step (f), we reorder the bonds of $A',B',C'$ and $D'$ for the next coarse graining.
 }
 \label{fig:skip_fga}
\end{figure}

As explained in \cite{ATRG}, 
the step (f), (g) and the step (a) of the next coarse graining
can be replaced by the following method 
which has no truncation errors and reduces the computational cost.
The flow of the improved method consistis of four steps which are
shown in step (c), (d), (e) and (f) of Fig.~\ref{fig:skip_fga}.
In step (c) of Fig.~\ref{fig:skip_fga}, 
$G$ and $H$ are decomposed in the coarse-grained direction as
\begin{align}
  G_{\Omega'_0x_0x_1} &= \sum_\gamma U_{\{G\}\Omega'_0x_0\gamma}S_{\{G\}\gamma\gamma}V_{\{G\} x_1 \gamma},\\
  H_{\Omega'_1x_1x_2} &= \sum_\delta U_{\{H\}\Omega'_1x_2\delta}S_{\{H\}\delta\delta}V_{\{H\} x_1\delta},
\end{align}
and $S_{\{G\}\gamma\gamma}, V_{\{G\} x_1 \gamma},
V_{\{H\} x_1\delta}$ and $S_{\{H\}\delta\delta}$ 
are contracted to obtain $K$,
\begin{align}
  K_{\gamma\delta} &= \sum_{x_1}
  S_{\{G\}\gamma\gamma}V_{\{G\} x_1 \gamma}
  V_{\{H\} x_1\delta}S_{\{H\}\delta\delta} \label{eq:make_K}.
\end{align}
In step (d) of Fig.~\ref{fig:skip_fga}, 
we decompose $K$ with the SVD,
\begin{align}
  K_{\gamma\delta} &= \sum_\alpha U_{\{K\}\gamma\alpha}S_{\{K\}\alpha\alpha}V_{\{K\}\delta\alpha},
\end{align}
and, in step (e) of Fig.~\ref{fig:skip_fga}, 
we contract the following tensors to make $A',B',C'$ and $D'$,
\begin{align}
  A'_{\Omega'_0x_0\alpha}&=\sum_{\gamma} U_{\{K\}\gamma\alpha} U_{\{G\}\Omega'_0x_0\gamma}\label{eq:A_dash},\\
  B'_{\Omega'_1 x_2\alpha}&=\sum_{\delta} S_{\{K\}\alpha\alpha}V_{\{K\}\delta\alpha}U_{\{H\}\Omega'_1x_2\delta}\label{eq:B_dash},\\
  C'_{\Omega'_0 x_0\alpha}&=\sum_{\gamma} S_{\{K\}\alpha\alpha}U_{\{K\}\gamma\alpha}U_{\{G\}\Omega'_0x_0\gamma}\label{eq:C_dash},\\
  D'_{\Omega'_1x_2\alpha}&=\sum_\delta V_{\{K\}\delta\alpha}U_{\{H\}\Omega'_1x_2\delta}\label{eq:D_dash}.
\end{align}
Finally, in step (f), we reorder the bonds of $A',B',C'$ and $D'$ for the next coarse graining
in a similar way to
the step (g) in Fig.~\ref{fig:atrg_algorithm}.

The total computational cost of the original ATRG is $O(\chi^{2d+1})$
and the total memory cost is $O(\chi^{2d})$.
The former cost comes from the step (e) in Fig.~\ref{fig:atrg_algorithm} and 
the swapping bonds part which consists of the step (b) and the step (c) in Fig.~\ref{fig:atrg_algorithm}.
The latter cost arises from the swapping bonds part.
Therefore, the swapping bonds part is the bottleneck one of 
both the computational cost and the memory cost in ATRG.

\section{Cost reduction of swapping bonds part\label{sec:cr_swapping}}
\begin{figure}[t]
  \centering
  \includegraphics[width=85mm]{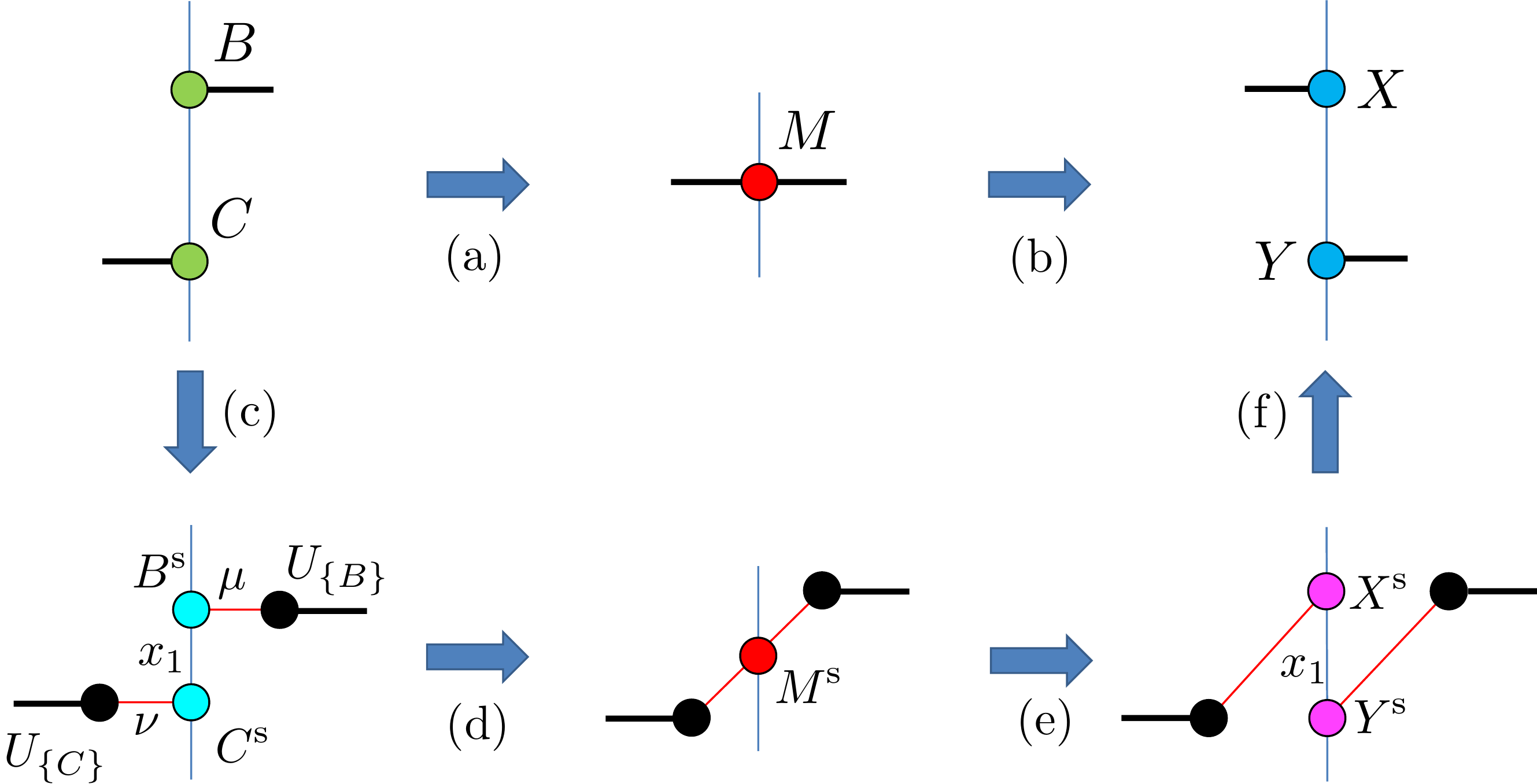}
  \caption{The cost reduction of the swapping bonds part in ATRG. 
    Step (a) and (b) represent the step (b) and (c) of Fig.~\ref{fig:atrg_algorithm}. 
    In step (c), we perform the SVDs of $B$ and $C$
    to make $B^{\mathrm{s}}$ and $C^{\mathrm{s}}$ 
    by the contractions in eq. (\ref{eq:cont_for_B^2}) and (\ref{eq:cont_for_C^2}).
    Note that there is no truncation in this step and
    the red lines mean that 
    the maximum bond dimensions of the indices, $\mu$ and $\nu$, 
    are taken to be $\chi^2$.
    In step (d), $M^{\mathrm{s}}$ is made of
    the contraction of $B^{\mathrm{s}}$ and $C^{\mathrm{s}}$ by summing up the bond $x_1$.
    In step (e), the combinations of the bonds are changed 
    from $\mu \alpha$ of $B^{\mathrm{s}}$ and $\beta \nu$ of $C^{\mathrm{s}}$
    to $\nu \alpha$ of $X^{\mathrm{s}}$ and $\mu \beta$ of $Y^{\mathrm{s}}$
    using the PSVD.
    In step (f), we contract 
    $X^{\mathrm{s}}$ and $U_{\{C\}}$ 
    by summing up the bonds $\nu$
    to make $X$,
    and contract 
    $Y^{\mathrm{s}}$ and $U_{\{B\}}$ 
    by summing up the bonds $\mu$
    to make $Y$.
  }
  \label{fig:cost_reduced_swapping}
\end{figure}

In this section we propose a swapping bonds method 
instead of the step (b) and (c) in Fig.~\ref{fig:atrg_algorithm}.
The flow of the proposed method is shown in step (c), (d), (e) and (f) 
of Fig.~\ref{fig:cost_reduced_swapping}.
The first step of the proposed swapping bonds method 
(step (c) of Fig.~\ref{fig:cost_reduced_swapping})
is the SVDs of $B$ and $C$ as
\begin{align}
  B_{\Omega_1 x_1\alpha}&= \sum_\mu U_{\{B\}\Omega_1\mu}S_{\{B\}\mu\mu}V_{\{B\} x_1\alpha \mu},\\
  C_{\Omega_2 x_1\beta}&= \sum_\nu U_{\{C\}\Omega_2\nu}S_{\{C\}\nu\nu}V_{\{C\} x_1\beta \nu},
\end{align}
and defining $B^{\mathrm{s}}$ and $C^{\mathrm{s}}$ as
\begin{align}
  B^{\mathrm{s}}_{x_1\alpha\mu}&= S_{\{B\}\mu\mu}V_{\{B\} x_1\alpha \mu} \label{eq:cont_for_B^2},\\
  C^{\mathrm{s}}_{x_1\beta\nu}&= S_{\{C\}\nu\nu}V_{\{C\} x_1\beta \nu} \label{eq:cont_for_C^2} .
\end{align}
Note that there is no truncation in this step and
the red lines in Fig.~\ref{fig:cost_reduced_swapping} 
mean that
the maximum bond dimensions of the indices, $\mu$ and $\nu$, 
are taken to be $\chi^2$.
In step (d) of Fig.~\ref{fig:cost_reduced_swapping}, 
we perform the contraction of $B^{\mathrm{s}}$ and $C^{\mathrm{s}}$
to make $M^{\mathrm{s}}$,
\begin{align}
  M^{\mathrm{s}}_{\mu\nu\alpha\beta} 
  &= \sum_{x_1}
  B^{\mathrm{s}}_{x_1\alpha\mu} C^{\mathrm{s}}_{x_1\beta\nu}.
\end{align}
Then, in step (e) of Fig.~\ref{fig:cost_reduced_swapping}, we change the combinations of the bonds 
from $\mu \alpha$ of $B^{\mathrm{s}}$ 
and $\beta \nu$ of $C^{\mathrm{s}}$ to 
$\nu \alpha$ of $X^{\mathrm{s}}$ and $\mu \beta$ of $Y^{\mathrm{s}}$ as
\begin{align}
  M^{\mathrm{s}}_{\mu\nu\alpha\beta} &\simeq \sum_{x_1=1}^\chi U_{\{M^{\mathrm{s}}\}\nu\alpha x_1}S_{\{M^{\mathrm{s}}\}x_1x_1}V_{\{M^{\mathrm{s}}\}\mu\beta x_1},\\
  X^{\mathrm{s}}_{\nu\alpha x_1} &= \sqrt{S_{\{M^{\mathrm{s}}\} x_1 x_1}} U_{\{M^{\mathrm{s}}\}\nu\alpha x_1},\\
  Y^{\mathrm{s}}_{\mu\beta x_1} &= \sqrt{S_{\{M^{\mathrm{s}}\} x_1 x_1}} V_{\{M^{\mathrm{s}}\}\mu\beta x_1},
\end{align}
using the PSVD.
Finally, in step (f) of Fig.~\ref{fig:cost_reduced_swapping}, 
we make $X$ and $Y$ as
\begin{align}
  X_{\Omega_2 \alpha x_1} &= \sum_{\nu} 
  X^{\mathrm{s}}_{\nu\alpha x_1} U_{\{C\}\Omega_2\nu},\\
  Y_{\Omega_1 \beta x_1} &= \sum_{\mu} 
  Y^{\mathrm{s}}_{\mu\beta x_1} U_{\{B\}\Omega_1\mu}.
\end{align}

By applying the above method to ATRG, 
we can reduce
the computational cost 
from $O(\chi^{2d+1})$ to $O(\chi^{\max(d+3,7)})$ in the swapping part
and the memory cost of the whole algorithm from
$O(\chi^{2d})$ to $O(\chi^{\max(d+1,6)})$
\footnote{Note that the memory cost can be reduced to $O(\chi^{\max(d+1,6)})$
if ${(\chi^{\mathrm{(init)}}})^{2d}< \chi^{\max(d+1,6)}$.}.

\section{Numerical results \label{sec:numerical_results}}
In this section, we present the results of the 
four-dimensional hypercubic Ising model 
with the periodic boundary condition
to examine the effectiveness of the proposed swapping bonds method.
We use the Ising model
which has $1024^4$ spin sites
and set the temperature to 6.68, which is 
close to the critical point of the Monte Carlo result \cite{MC_4d-ising}.
We use the RSVD as the PSVD part,
and the parameters of the RSVD iterations and oversamples
are taken to be $\chi$ and $2\chi$ respectively.
The detail study of these parameters is given in appendix~\ref{sec:rsvd}.

\begin{figure}[t]
\centering
\includegraphics[width=85mm]{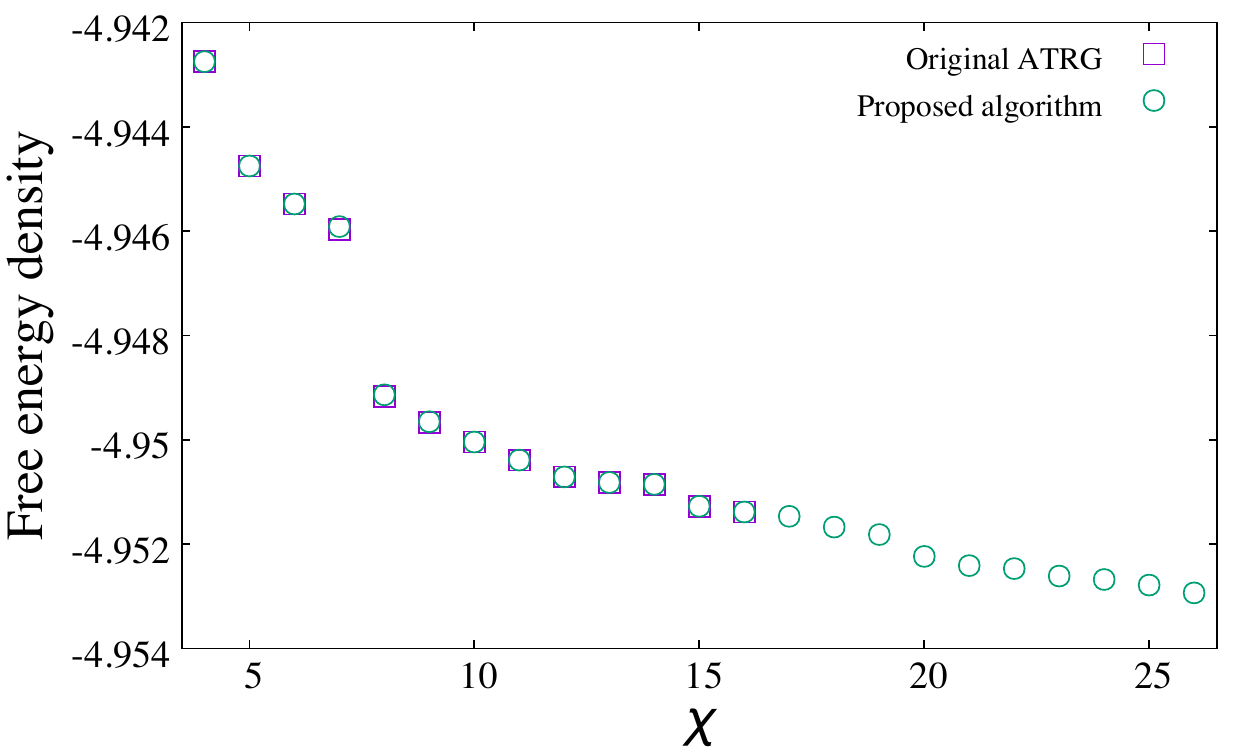}
 \caption{
   The free energy density of the four dimensional Ising model as 
   a function of $\chi$ on $1024^4$ lattice.
   The temperature is set to 6.68.
   The result of the proposed algorithm is consistent with that of 
   the original ATRG.
 }
 \label{fig:f_chi_atrg_proposed}
\end{figure}

\begin{figure}[t]
\centering
\includegraphics[width=85mm]{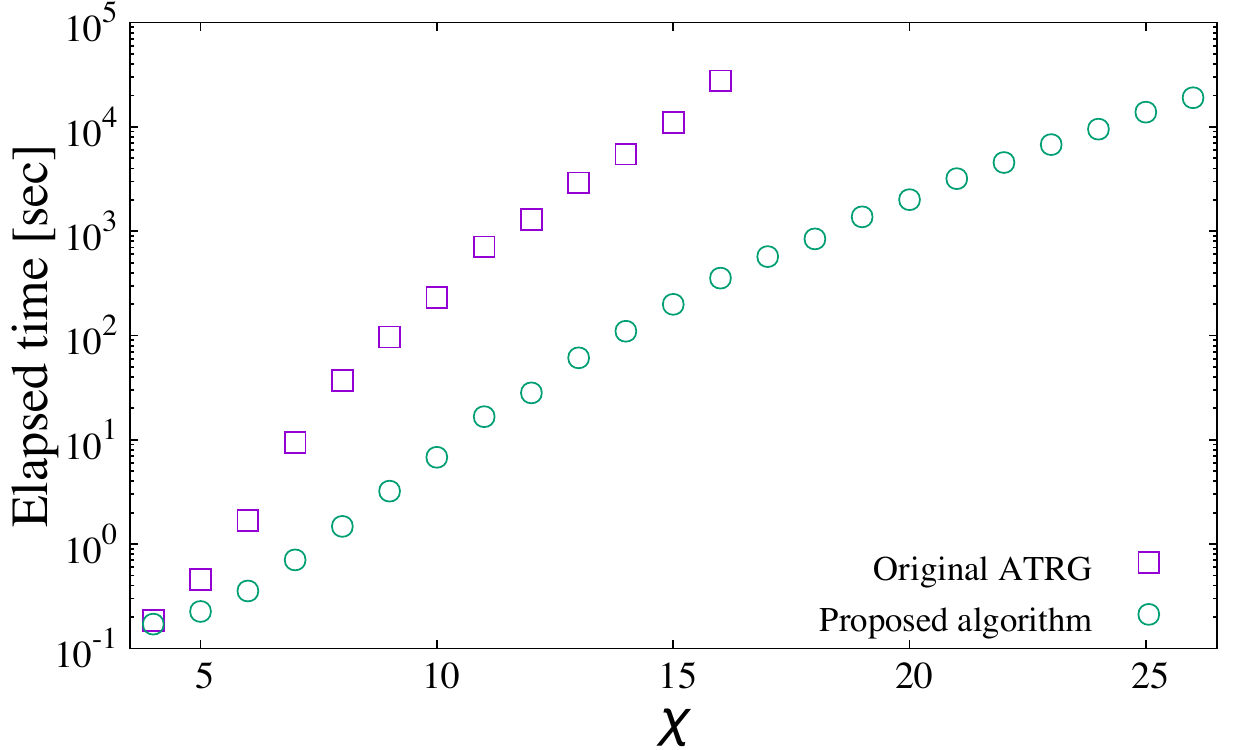}
\caption{
  The total elapsed time as a function of $\chi$. 
  The time of the proposed algorithm is significantly reduced from that of the original ATRG for larger $\chi$. 
}
\label{fig:time_chi_atrg_proposed}
\end{figure}

\begin{figure}[t]
\includegraphics[width=85mm]{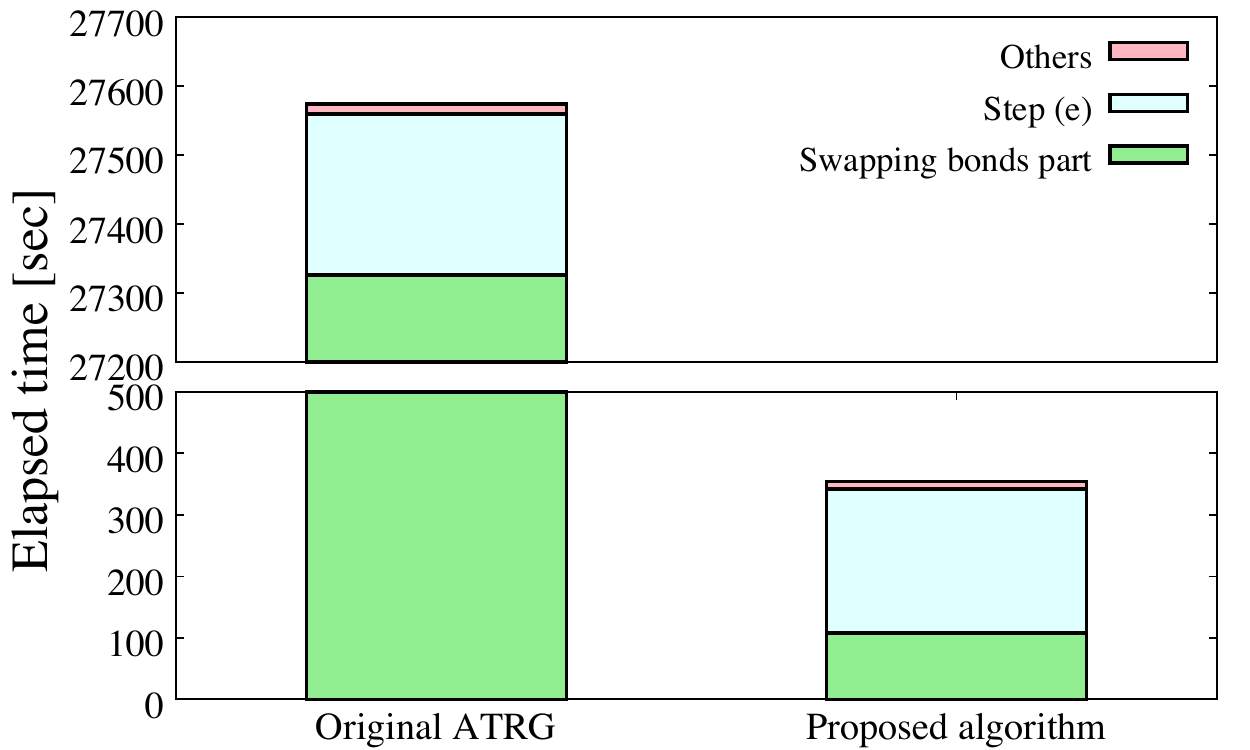}
 \caption{The detail of the total elapsed time at $\chi=16$.
   The bottleneck part in the original ATRG is the swapping bonds part,
   while the step (e) in Fig.~\ref{fig:atrg_algorithm}
   turns out to be the most time-consuming part in the proposed algorighm.
 }
 \label{fig:detail_time_chi_16}
\end{figure}

First, we show the comparison between ATRG using the proposed method 
and the original ATRG.
The free energy densities obtained by the two algorithms
are shown in
Fig.~\ref{fig:f_chi_atrg_proposed}.
From this figure,
we can see that the result of the proposed algorithm is consistent 
with that of the original ATRG.
This consistency would be 
because the rank of $M$ in the original ATRG is $\chi^3$
and coincides with the rank of $M^{\mathrm{s}}$ in the proposed algorithm.
Next, Fig.~\ref{fig:time_chi_atrg_proposed} shows the total elapsed times of 
the two algorithms
and the time of the proposed algorithm is significantly reduced from that of the original ATRG for larger $\chi$. 
From Fig.~\ref{fig:f_chi_atrg_proposed} and~\ref{fig:time_chi_atrg_proposed},
we conclude that 
the performance of ATRG can be improved
without the loss of the accuracy of the free energy density.
Figure~\ref{fig:detail_time_chi_16} shows 
the details of the elapsed times of the two algorithms at $\chi=16$.
We can see that 
the speeding up of ATRG 
is attained by the cost reduction of the swapping bonds part
and the step (e) in Fig.~\ref{fig:atrg_algorithm},
which contains $O(\chi^9)$ contractions,
turns out to be the most time-consuming part
in the proposed algorithm.

\begin{figure}[t]
\centering
\includegraphics[width=85mm]{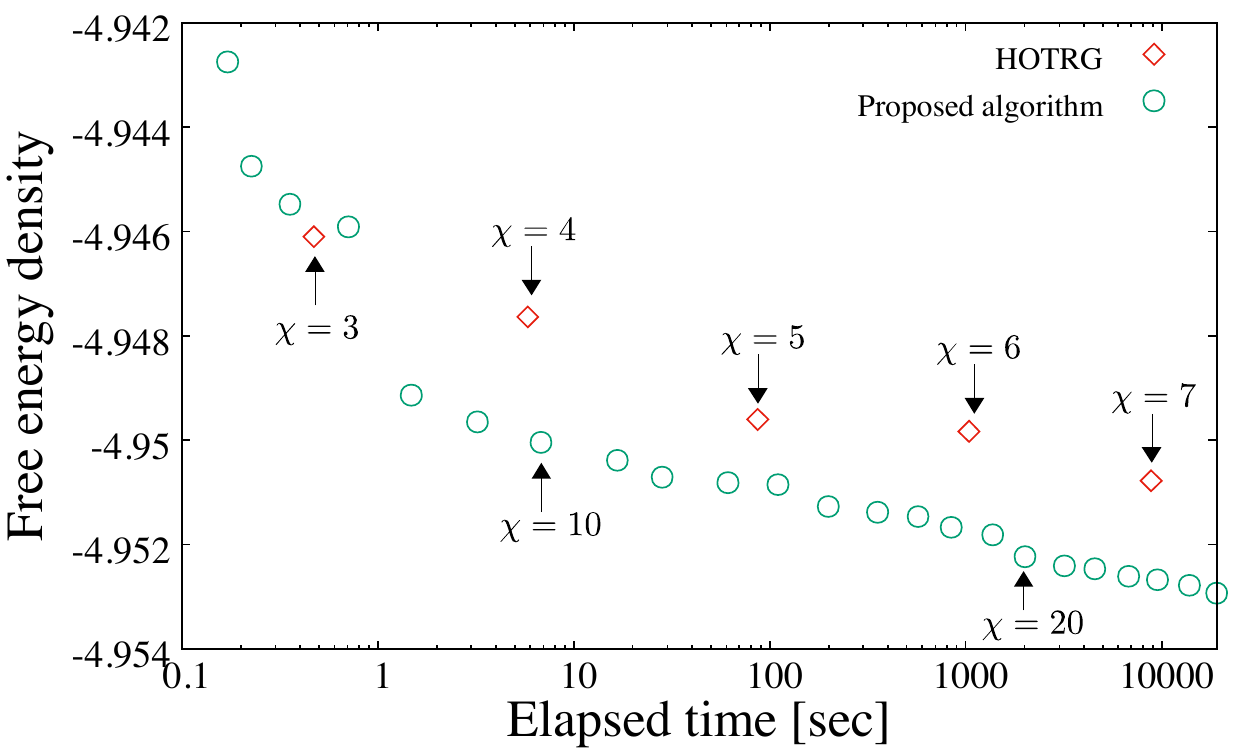}
\caption{
  The free energy density of the four dimensional Ising model as 
  a function of the elapsed time on $1024^4$ lattice and 
  the temperature is set to 6.68.
  The result of the proposed algorithm 
  can reach lower than that of HOTRG 
  in the same elapsed time
  when the maximum bond dimensions of the two algorithms are increasing.
}
 \label{fig:f_time_hotrg_proposed}
\end{figure}

Next, we compare the results of the proposed algorithm and HOTRG.
The maximum bond dimensions in HOTRG 
are set to the integers in $3\leq \chi \leq 7$.
Figure~\ref{fig:f_time_hotrg_proposed} shows 
the relation between the free energy densities
and the elapsed times of the two algorithms.
The elapsed times are monotonically increasing when the bond dimensions are growing
in this figure.
We assume that the larger the maximum bond dimension $\chi$ is, 
the higher the accuracy of the free energy density becomes.
Since the free energy densities of the proposed algorithm and HOTRG are monotonically decreasing
as the maximum bond dimensions are increasing in Fig.~\ref{fig:f_time_hotrg_proposed}, 
we can consider that
the lower the free energy density is, the higher the accuracy of that becomes.
From this view point, 
the proposed algorithm can obtain
more accurate free energy density than 
HOTRG in the same elapsed time
when the maximum bond dimensions of the two algorithms are increasing.

We use the machine 
which has 132GB for the memory 
and Intel(R) Xeon(R) CPU E5-2690 v4 @ 2.60GHz for the CPU.
The programs of the algorithms are written with python 3.7.1 and 
we use numpy.tensordot in numpy 1.15.4 for the tensor contractions and 
scipy.linalg.svd in scipy 1.1.0 for the SVD of the tensors and 
sklearn.utils.extmath.randomized\_svd in scikit-learn 0.20.1 for the RSVD of the tensors.

\section{Summary\label{sec:summary}}
We propose the swapping bonds method of which 
the computational cost can be reduced from $O(\chi^{2d+1})$ to $O(\chi^{\max(d+3,7)})$.
Moreover, the memory cost of the whole algorithm can be reduced from $O(\chi^{2d})$ to $O(\chi^{\max(d+1,6)})$ with the method. 
We examine ATRG using the method with the four-dimensional hypercubic Ising model and find that
the free energy density obtained by the algorithm is consistent 
with that obtained by the original ATRG while the elapsed time is significantly reduced.
Moreover, we compare the proposed algorithm with HOTRG 
in terms of the free energy density of the Ising model and the elapsed time.
Then, we find that
the proposed algorithm can obtain
more accurate free energy density than 
HOTRG in the same elapsed time
for the large maximum bond dimensions.

For the future works, 
the step (e) in Fig.~\ref{fig:atrg_algorithm},
which has the $O(\chi^{2d+1})$ contractions,
is the bottleneck part of the proposed algorithm.
To speed up this part, we will apply the paralell computings to the bottleneck contractions.
Furthermore, we will analyze physical quantities of the four-dimensional Ising model with large maximum bond dimensions
to compare with other results \cite{MC_4d-ising,4D-HOTRG}.

\begin{acknowledgments}
The author would like to thank Shinji Takeda for 
insightful feedbacks.

\end{acknowledgments}

\appendix
\section{Making $E$ and $F$\label{sec:making_projectors}}
\begin{figure}[htbp]
\centering
\includegraphics[width=65mm]{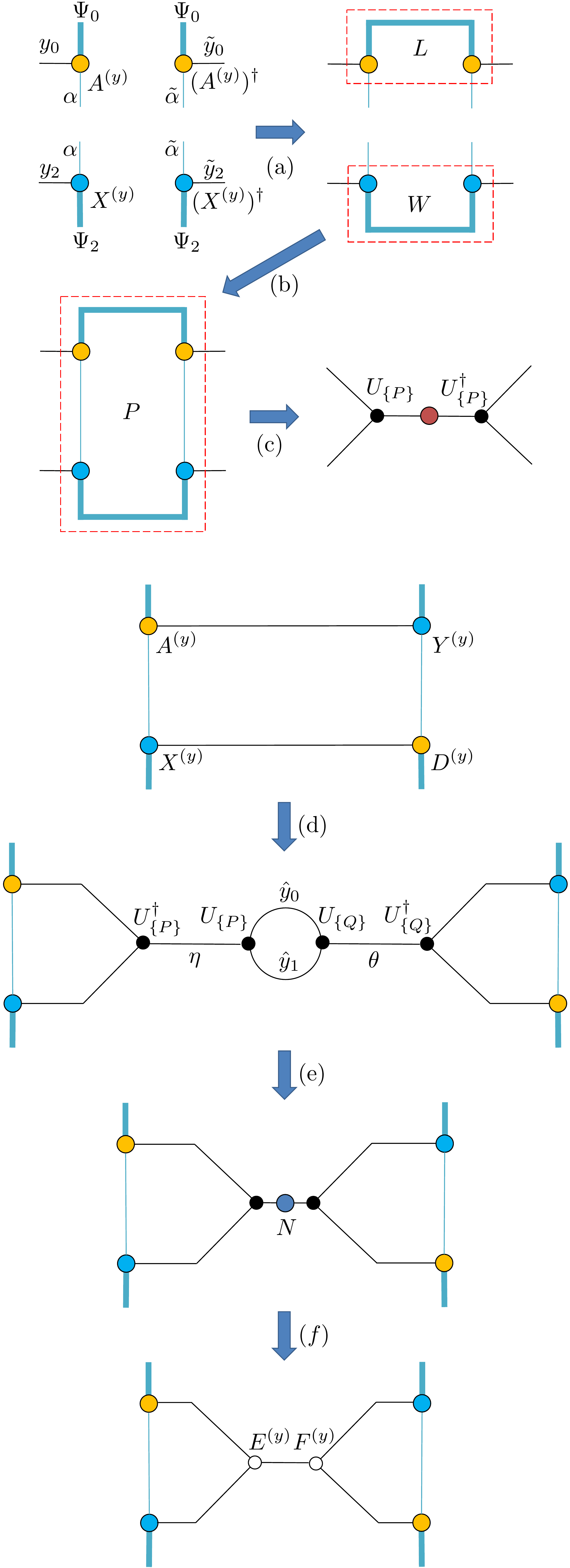}
 \caption{The flow of the making $E$ and $F$ for $y$-axis in step (d) of Fig.~\ref{fig:atrg_algorithm}.
   In step (a), we contract a tensor, $A^{(y)}(X^{(y)})$, and its conjugate 
   to make a tensor, $L(W)$.
   In step (b), $L$ and $W$ are contracted by summing up the bonds $\alpha$ and $\tilde{\alpha}$.
   In step (c), we perform the SVD or the eigenvalue decomposition of $P$ and 
   the truncation to retain $\chi$ largest singular values.
   The step (a), (b) and (c) are the example of obtaining $U_{\{P\}}$
   and we can also obtain $Q$ and $U_{\{Q\}}$ from $Y$ and $D$ similarly.
   In step (d), the isometries, $U^\dag_{\{P\}},U_{\{P\}},U_{\{Q\}}$ and $U^\dag_{\{Q\}}$,
   are inserted between $A^{(y)}X^{(y)}$ and $Y^{(y)}D^{(y)}$.
   In step (e), $U_{\{P\}}$ and $U_{\{Q\}}$ are contracted by summing up the bonds $\hat{y}_0$ and $\hat{y}_1$ to make $N$.
   In step (f), we decompose $N$ with the SVD and make $E^{(y)}$ and $F^{(y)}$ 
   by the contractions in eq. (\ref{eq:E^y}) and (\ref{eq:F^y}) respectively.
 }
 \label{fig:making_projectors}
\end{figure}

In this appendix, 
we explain a method of making the tensors, $E$ and $F$,
in step (d) of 
Fig.~\ref{fig:atrg_algorithm}.
Figure~\ref{fig:making_projectors}
shows the flow of the method to make
the tensors for $y$-axis.
Before starting the method for $y$-axis,
we reorder the bonds of $A$ and $X$ as 
\begin{align}
  A^{(y)}_{\Psi_0 y_0\alpha} &= A_{\Omega_0 x_0\alpha},\\
  X^{(y)}_{\Psi_2 \alpha y_2} &= X_{\Omega_2 \alpha x_1}, 
\end{align}
where the superscript $(y)$ represents the $y$-axis direction
and $\Psi_0=(x_0,z_0,w_0)$ and $\Psi_2=(x_1,z_2,w_2)$.
In step (a) of Fig.~\ref{fig:making_projectors}, 
the tensors and their conjugate tensors are contracted 
by summing up $\Psi$s as 
\begin{align}
  L_{y_0\alpha \tilde{y}_0\tilde{\alpha}} &= \sum_{\Psi_0} A^{(y)}_{\Psi_0 y_0\alpha}(A^{(y)}_{\Psi_0 \tilde{y}_0\tilde{\alpha}})^*,\\
  W_{y_2\alpha \tilde{y}_2\tilde{\alpha}} &= \sum_{\Psi_2} X^{(y)}_{\Psi_2 \alpha y_2}(X^{(y)}_{\Psi_2 \tilde{\alpha} \tilde{y}_2})^*, 
\end{align}
In step (b) of Fig.~\ref{fig:making_projectors}, we contract $L$ and $W$ 
by summing up the bonds $\alpha$ and $\tilde{\alpha}$,
\begin{align}
  P_{y_0 y_2 \tilde{y}_0 \tilde{y}_2} &= \sum_{\alpha,\tilde{\alpha}} L_{y_0\alpha \tilde{y}_0\tilde{\alpha}}W_{y_2\alpha \tilde{y}_2\tilde{\alpha}}. 
\end{align}
In step (c) of Fig.~\ref{fig:making_projectors}, 
we perform the SVD or the eigenvalue decomposition of $P$ and the truncation to retain $\chi$ largest singular values,  
\begin{align}
  P_{y_0 y_2 \tilde{y}_0 \tilde{y}_2} &\simeq \sum_{\eta=1}^\chi U_{\{P\}y_0 y_2\eta}S_{\{P\}\eta\eta}U^*_{\{P\} \tilde{y}_0 \tilde{y}_2\eta}, 
\end{align}
where $U_{\{P\}}$ is an isometry made by the left singular vectors which corresponds to the $\chi$ largest singular values.
We can also obtain $Q$ and $U_{\{Q\}}$ from $Y$ and $D$ in a similar way to $P$ and $U_{\{P\}}$.
Then, in step (d) of Fig.~\ref{fig:making_projectors},
the isometries, $U^\dag_{\{P\}},U_{\{P\}},U_{\{Q\}}$ and $U^\dag_{\{Q\}}$,
are inserted between $A^{(y)}X^{(y)}$ and $Y^{(y)}D^{(y)}$.
To make $E^{(y)}$ and $F^{(y)}$ explicitly, 
$U_{\{P\}}$ and $U_{\{Q\}}$ are contracted as
\begin{align}
  N_{\eta \theta} &= \sum_{\hat{y}_0,\hat{y}_1} U_{\{P\}\hat{y}_0\hat{y}_1\eta}U_{\{Q\}\hat{y}_0\hat{y}_1\theta}.
\end{align}
in step (e) of Fig.~\ref{fig:making_projectors}.
Next, in step (f) of Fig.~\ref{fig:making_projectors}, 
we carry out the SVD of $N$,
\begin{align}
  N_{\eta \theta} &= 
  \sum_\kappa U_{\{N\} \eta \kappa}S_{\{N\} \kappa\kappa}V_{\{N\} \theta\kappa},
\end{align}
and the contractions for making $E^{(y)}$ and $F^{(y)}$ as
\begin{align}
  E^{(y)}_{y_0y_2\kappa} &= 
  \sum_{\eta} \sqrt{S_{\{N\} \kappa\kappa}}U_{\{N\} \eta \kappa}U^*_{\{P\}\eta y_0 y_2},
  \label{eq:E^y}\\
  F^{(y)}_{y_1y_3\kappa} &= 
  \sum_{\theta} \sqrt{S_{\{N\} \kappa\kappa}}V_{\{N\} \theta\kappa} U^*_{\{Q\}\theta y_1 y_3}.
  \label{eq:F^y}
\end{align}
We can use the same way to another direction,
and 
$E$ and $F$ consist of 
the tensors of the directions
other than the coarse-graining one such as
\begin{align}
  E&=E^{(y)}\otimes E^{(z)}\otimes E^{(w)}\\
  F&=F^{(y)}\otimes F^{(z)}\otimes F^{(w)}.
\end{align}

To complete the making $E$ and $F$ step, it needs $4(d-1)$ times $O(\chi^{d+3})$ contractions,
$2(d-1)$ times $O(\chi^6)$ contractions 
and $2(d-1)$ times $O(\chi^6)$ SVDs
\footnote{The costs of these SVDs can be reduced to $O(\chi^5)$ if one uses the PSVDs.}.
Then, the bottleneck of this part is the $O(\chi^{d+3})$ contraction.

\section{Study of RSVD parameters\label{sec:rsvd}}
In this appendix, we study the parameters of the RSVD 
in the proposed algorithm and the original ATRG.
The purpose of this study is to see
how large the parameters of each algorithm should be taken
to converge to the free energy density obtained by an ATRG algorithm 
which uses the truncation of the full SVD instead of the RSVD.

\begin{figure}[t]
  \centering
  \includegraphics[width=85mm]{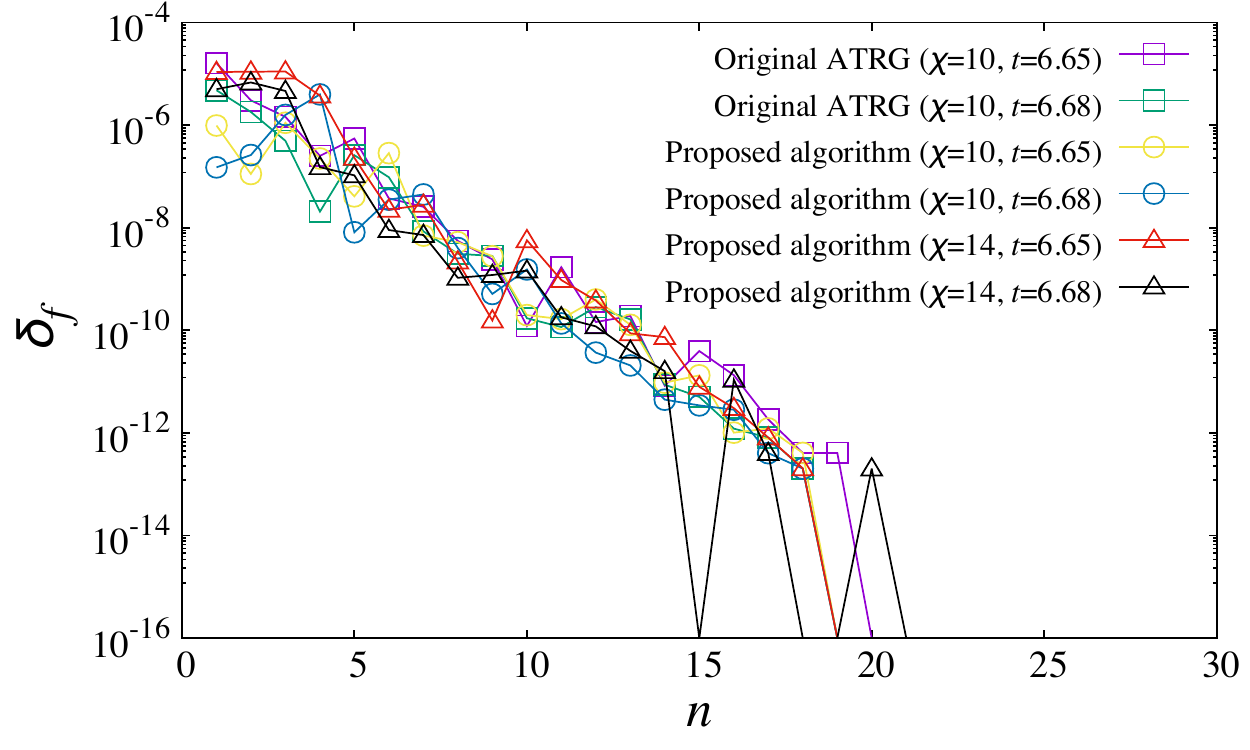}
  \caption{The relative errors of the free energy densities of 
    the Ising model on $1024^4$ lattice
    with the original ATRG and the proposed algorithm,
    where 
    $\chi=10,14$,
    the temperatures are set to 6.65 and 6.68 ($t=6.65, 6.68$), 
    and then the number of the RSVD iterations is taken to be $\chi$.
    The oversampling parameter $n$ is changed from 1 to $2\chi$.
    We find that 
    the results of the original ATRG and 
    the proposed algorithm show similar behavior at $\chi=10$
    and 
    it is large enough to take $n\geq 20$ 
    at $\chi=10,14$.
  }
  \label{fig:df_oversamples}
\end{figure}

To investigate the convergence
for the original ATRG,
we use the Ising model on the $1024^4$ lattice 
and define the relative error of the free energy density as
\begin{align}
  \delta_f =\left|\frac{f_{n,\mathrm{RSVD}}-f_{\mathrm{SVD}}}{f_{\mathrm{SVD}}}\right|, \label{eq:relative_error}
\end{align}
where
$f_{\mathrm{SVD}}$ is the free enegy density 
obtained by 
ATRG
which applies the truncation of the full SVD to $M$ and
$f_{n,\mathrm{RSVD}}$ is the one using the original ATRG which applies the RSVD of 
$n$ oversamples and $\chi$ iterations to the matrix.
In the case of the proposed algorithm, 
we can also define $\delta_f$ similarly.
Note that 
if the number of the RSVD iterations is fixed to $\chi$,
the computational cost of the swapping bond part in the original ATRG becomes 
$O(\chi^{10})$, and then the leading cost of the algorithm gets worse from $O(\chi^9)$.
On the other hand, 
the cost of the part in the proposed algorithm becomes $O(\chi^8)$
and it is not the leading one in the whole algorithm
(see Fig.~\ref{fig:detail_time_chi_16}).
Figure~\ref{fig:df_oversamples} shows $\delta_f$ 
as the function of $n$
with $\chi=10,14$ and 
the temperatures are taken to be 6.65 and 6.68 ($t=6.65, 6.68$).
From Fig.~\ref{fig:df_oversamples}, 
we can see that 
the results of the original ATRG and 
the proposed algorithm show similar behavior at $\chi=10$ 
and it is large enough to take 
$n\geq 20$
at $\chi=10,14$.
Therefore,
we use $\chi$ and $2\chi$ for the numbers of the RSVD iterations
and oversamples respectively
for the numerical results shown in Sec.~\ref{sec:numerical_results}.
We believe that it is large enough for the RSVD 
to converge to the full SVD results 
when $\chi\geq 10$.
Finally we remark that
when 
degenerate singular values are truncated,
the results of the RSVD may not converge to those of the full SVD 
even if the parameters are taken to be sufficiently large.

\end{document}